\documentclass[12pt,preprint]{aastex6}

\shorttitle{the time evolution of HH~2}
\shortauthors{Raga et al.}

\begin{document}

\title{The time evolution of HH~2 from four epochs of HST images}

\author{A. C. Raga\altaffilmark{1}, B. Reipurth\altaffilmark{2},
P. F. Vel\'azquez\altaffilmark{1}, A. Esquivel\altaffilmark{1}, J. Bally\altaffilmark{3}}
\altaffiltext{1}{Instituto de Ciencias Nucleares, Universidad
Nacional Aut\'onoma de M\'exico, Ap. 70-543, 04510 D.F., M\'exico}
\altaffiltext{2}{Institute for Astronomy, University of Hawaii at Manoa, Hilo, HI 96720, USA}
\altaffiltext{3}{Center for Astrophysics and Space Astronomy, University of Colorado, UCB 389,
Boulder, CO 80309, USA}

\email{raga@nucleares.unam.mx}

\begin{abstract}
  We have analyzed four epochs of H$\alpha$ and [S~II] HST images of the HH~1/2 outflow
  (covering a time interval from 1994 to 2014)
  to determine proper motions and emission line fluxes of the knots of HH~2. We find
  that our new proper motions agree surprisingly well with the motions measured by
  Herbig \& Jones (1981), although there is partial evidence for a slight
  deceleration of the motion of the HH~2 knots from 1945 to 2014. We also measure
  the time-variability of the H$\alpha$ intensities and the [S~II]/H$\alpha$ line ratios,
  and find that knots H and A have the largest intensity variabilities (in $1994\to 2014$).
  Knot H (which now dominates the HH~2 emission) has strengthened substantially, while
  keeping an approximately constant [S~II]/H$\alpha$ ratio. Knot A has dramatically faded,
  and at the same time has had a substantial increase in its [S~II]/H$\alpha$ ratio.
  Possible interpretations of these results are discussed.
\end{abstract}

\keywords{shock waves --- stars: winds, outflows ---
Herbig-Haro objects --- ISM: jets and outflows ---
ISM: individual objects (HH2)}

\section{Introduction}

The HH~1/2 system is probably the best studied Herbig-Haro flow from a young star (see the
review of Raga et al. 2011). This object consists of a central source seen at
radio wavelengths (Pravdo et al. 1985; Rodr\'\i guez et al. 2000), a bipolar
jet/counter jet system (seen in the IR, Noriega-Crespo \& Raga 2012), and two
``heads'' (HH~1 to the NW and HH~2 to the SE) at approximately $1'$ from the outflow
source. The HH~1/2 outflow axis lies within $\sim 5^\circ$ from the plane of the sky (see,
e.~g., B\"ohm \& Solf 1985).

There are now four epochs of Hubble Space Telescope (HST) images of the HH~1/2 system, covering a
$\sim 20$~yr time span. These images were used by Raga et al. (2016a) to measure the proper motions
and the time-evolution of the line intensities of HH~1. In the present paper we
use the same set of images to study the time-evolution of HH~2.

While HH~1 is a relatively compact collection of knots (with an angular
size of $\sim 15''$), showing a general organization
into a bow-like structure, HH~2 is angularly more extended ($\sim 30''$),
and has a relatively chaotic condensation structure (see, e.g., Raga et al.
2015b). Because of this, it is more complex to untangle the time-evolution of this object.

In this paper, we calculate the proper motions of the HH~2 condensations. This
represents a continuation of the work of Herbig \& Jones (1981), Eisl\"offel et al. (1994)
and Bally et al. (2002), who have previously studied the motions of HH~2.

We also calculate the time-evolution of the H$\alpha$ intensities and the [S~II]/H$\alpha$ ratios
of the HH~2 condensations. This is to some extent an extension of the work of Herbig (1969),
who studied the time-evolution of the broad-band emission of of the HH~2 knots, and
of the work of Raga et al. (2016b), who compared the angularly integrated emission of
HH~2 obtained from HST images and the spectrophotometric observations of Brugel
et al. (1981).

The paper is organized as follows. In Section 2 we describe the available set of four
epochs of HST images of HH~2. In Section 3, we describe the qualitative time-evolution of
the morphology of HH~2. In Section 4, we present the proper motion determinations. In
Section 5, we discuss the time evolution of the H$\alpha$ intensity and [S~II]/H$\alpha$
line ratios of the HH~2 condensations. Section 6 discusses possible interpretations
of the motions and intensity variabilities of the HH~2 knots. Finally, the results are
summarized in Section 7.

\section{The observations}

There are now four epochs of HST H$\alpha$ images (obtained with the F656N filter) and red [S~II] images
(with the F673N filter) of the HH~1/2 outflow, which we label:
\begin{itemize}
\item 1994: the 1994.61 images of Hester et al. (1998),
\item 1997: the 1997.58 images of Bally et al. (2002),
\item 2007: the 2007.63 images of Hartigan et al. (2011),
\item 2014: the 2014.63 images of Raga et al. (2015a, b).
\end{itemize}
The first three epochs were obtained with the WFPC2 camera, and the fourth epoch with
the WFC3. The F656N and F673N filters in of the two cameras are roughly equivalent.
A description of the characteristics of these images is given by Raga et al. (2016a), who used
them to study the time-evolution of HH~1. In the present paper we use the same images (not corrected
for reddening), with the flux calibration and the alignment (between the successive
epochs) described by Raga et al. (2016a). All of the images have a $0''.1$ per pixel scale.

 { As discussed
  by Raga et al. (2015a, 2016a), the H$\alpha$ filters include a contribution of the [N~II]~6548
  line at a level of $\sim 2$\% and a continuum flux at a level of $\sim 1$\%\  (of the H$\alpha$
  flux). The [S~II] filters are contaminated by the continuum emission at a level
  of $\sim 1$\%\ of the [S~II] flux. As the initial images have a good signal-to-noise ratio
  and because additionally our flux measurements are done in decreased angular
  resolution images (which were convolved with a $1''$ wavelet, see section 5), the errors due
  to photon statistics are small compared to the line and continuum contamination of the
  filters.

  The main contribution to the errors in the line fluxes lies in the flux calibration of the filters.
  As described by Raga et al. (2016a), the standard continuum calibration
  of the filters (see, e.g., the WFC3 Instrument Handbook) has been converted into a calibration
  for the line fluxes using the full width of the filters (see the discussion of O'Dell et al. 2013).
  This gives an ``exact'' calibration only for a ``box car'' filter transmission
  function, but of course has errors that depend on the exact position within the filter of
  the emission line for the actual curved filter transmissions. The errors associated with
  this effect can be estimated by calculating the standard deviation from a box car function
  of the actual filter transmission curves (given in the WFPC2 and
  WFC3 Instrument Handbook). From the plotted curves, we obtain standard deviations
  of $(8.5,3.2)$\%\ (as a percentage of the mean transmission)
  for the F656N filter and $(6.1,4.8)$\%\ for the F673N filter (the first
  of the values within the parentheses corresponding to the WFPC2 camera filters and the second
  one to the WFC3 camera).

  We then estimate the errors in the H$\alpha$ and [S~II] line fluxes considering the line/continuum contamination
  and the flux calibration errors. In this way we obtain estimates of $(9,4)$\%\ for
  the line flux errors of the F656N frames (as percentages of the H$\alpha$ fluxes)
  and $(6,5)$\%\ for the F673N filters (as percentages of the [S~II] line fluxes), with
  the first and second values in the parentheses corresponding to the WFPC2 and WFC3 cameras, respectively.}

{ The four epochs of H$\alpha$ images are shown in Figure 1 (with a field that includes all
of the HH~2 emission) and in Figure 2 (a smaller field including the brighter knots H and A),
and the four [S~II] images are shown in Figures 3 and 4. Figures 5 and 6 show the four
H$\alpha$ and [S~II] images (respectively), convolved with a ``Mexican hat'' wavelet with
a central peak of $1''$ radius. These convolved images have a resolution comparable
to ground-based images.}

We note a particular difficulty that is found when measuring the positions of the knots of HH~2.
The images are scaled, rotated and aligned using the positions of the Cohen-Schwartz star
(Cohen \& Schwartz 1979) and ``star no. 4'' of Str\"om et al. (1985). The positions of these
two stars (the CS star along the outflow axis and star no. 4 approximately $1'$ to the E
of the CS star) are closer by a factor of $\sim 3$ to HH~1 than to HH~2. Because of this,
the error in the positions of the HH~2 knots (dominated by the effect of the rotation
of the frames rather than by the scaling) are considerably larger than the errors for
the positions of HH~1 (measured by Raga et al. 2016a). This problem in measuring the
positions of the HH~2 knots in CCD frames that only include these two stars was
already noted in the first attempt to carry out astrographic measurements of HH objects
with CCD frames (Raga et al. 1990).

We find that the errors in the centering/scaling/rotation of the four epochs of
images of HH~2 are dominated by offsets perpendicular to the outflow axis (with
a PA=325$^\circ$ direction see Bally et al. 2002), with values of $\sim 0''.2$.
These errors result in systematic shifts (between the successive epochs)
of all of the HH~2 knots in the direction perpendicular to the outflow axis.
One could in principle apply a small correction to the rotations of the successive
frames in order to minimize these systematic offsets, but we feel that it is
better not to apply such a correction.

\section{Morphological changes of HH~2}

In Figure 1, we show the H$\alpha$ emission of HH~2 in the four epochs. In the 1994
frame (top left frame), we show the identifications of the emitting knots of Herbig \& Jones (1981) and
Eisl\"offel et al. (1994). It is clear that while the knots in the periphery of HH~2 (knots T, E, L
and D) have a similar morphology in the four frames, the knots in the brighter, central region
of HH~2 (knots G, H and A) have rather dramatic changes. Knot K is a peripheral knot, and it
shows considerable intensity changes while preserving its general morphology.

The time-evolution of part of the central region of HH~2 (i.e., the region shown with a white box
in the bottom right frame of Figure 1) is shown in Figure 2. In this figure, we see that condensation
H has dramatic morphological changes (as well as a strong brightening) and condensation A an
also dramatic fading during the period covered by the four HST frames.

The four [S~II] frames are shown in Figure 3. We find a behaviour similar to the H$\alpha$
images, with the knots on the periphery of HH~2 preserving their morphological characteristics,
and with large variabilities in the central knots. This effect is clearly seen in Figure 4,
which shows the [S~II] emission of the central region of HH~2.

\section{The proper motions}

Given the large morphological changes in the HH~2 knots at subarcsecond scales (see Figures 2 and 4),
in order to obtain proper motions, following Raga et al. (2016a) we first convolve the images with
a ``Mexican hat'' wavelet with a central peak of radius $\sigma=1''$ (see equation 1 of Raga et al. 2016a).
The resulting H$\alpha$ (see Figure 5) and [S~II] (see Figure 6) convolved images show well defined
peaks for the condensations L, E, G, H, I, T, A and D. In the convolved images, Knot K has two peaks
which we call K1 (the E peak) and K2 (the W peak, see Figures 5 and 6).

We then carry out paraboloidal fits to
$3\times 3$ pixel regions centered on these peaks and measure offsets with respect to the knot
positions in the 1994 frame. We adopt a PA=145$^\circ$
position angle for the outflow axis of HH~2
(with a direction opposite to the PA=325$^\circ$ direction of the HH~1 jet obtained by
Bally et al. 2002) and calculate the offsets parallel ($\Delta x$) and perpendicular ($\Delta y$)
to the outflow axis (with positive values of $\Delta y$ directed to the W of the outflow axis)
between the knots in the successive frames and their positions
in the 1994 frame. The offsets obtained in the H$\alpha$ and in the [S~II]
frames are shown in Figures 7 and 8.

In Figures 7 and 8 we see that:
\begin{itemize}
\item the H$\alpha$ and [S~II] offsets are consistent for all knots, except for
  the offsets along the outflow axis ($\Delta x$) of knot I and the $\Delta y$ offsets of knot L,
\item the largest offsets along the outflow axis ($\Delta x$) are seen for
  knot I (with inconsistent H$\alpha$ and [S~II] offsets), and are therefore somewhat suspect,
\item the central knots of HH~2 (knots G, H and A) have $\Delta x\sim 5 \Delta y$ and the peripheral
  knots (knots E, K1, K2, T and D) have $\Delta x \sim \Delta y$,
\item a systematic ``down/up'' pattern (with an amplitude of $\sim 0''.2$)
  is seen in the successive $\Delta y$ values, which
  probably results from the frame centering problems discussed in Section 2.
\end{itemize}

Table 1 shows the velocities parallel ($v_x$) and perpendicular ($v_y$) to the outflow axis
obtained from linear, least squares fits to the H$\alpha$ and [S~II] offsets and assuming
a distance of 414~pc to HH~2 (Menten et al. 2007). The
$\epsilon(v_x)$ and $\epsilon(v_y)$ errors computed from the fits are also given.
The linear fits are plotted for all of the knots in Figures 7 and 8,
showing that the scatter of the measured points is large enough that a fit with a higher
order polynomial (as would be needed to estimate an acceleration/deceleration for the knots)
would be meaningless.

In Table 1, we also give the proper motion velocity $v_t=\sqrt{v_x^2+v_y^2}$ and its associated error
$\epsilon(v_t)$, as well as the proper motion velocities $v_{HJ}$ obtained by Herbig \& Jones (1981)
for some of the knots. The values of $v_{HJ}$ were obtained from the analysis of red photographic
plates obtained from 1945 to 1980 of Herbig \& Jones (1981), which we have now corrected to
a distance of 414~pc to HH~2.

The consistency between our values of $v_t$ and the corresponding $v_{HJ}$ values of
Herbig \& Jones (1981) is quite remarkable:
\begin{itemize}
\item for knots E, G, H and D we obtain slightly lower proper motion velocities than
  the $v_{HJ}$ values of Herbig \& Jones (1981), though the difference in velocities is probably only
  significant for knots G and D,
\item for knot A, we obtain a higher proper motion velocity than its $v_{HJ}$ value, but
  this is probably not a significant result given the problems found by Herbig \& Jones (1981)
  for determining the proper motion of this condensation (for which they did not give
  an estimated error),
\item for knot I (with inconsistent H$\alpha$ and [S~II] proper motions, see above),
  we see that our H$\alpha$ proper motion velocity is somewhat higher than $v_{HJ}$
  and that our [S~II] velocity is substantially lower.
\end{itemize}
Therefore, what can be said is that for some of the knots (knots E, G, H and D)
we might be seeing a slight slowing down between the $1945\to 1980$ period (analyzed by
Herbig \& Jones 1981) and the $1994\to 2014$ period that we are now analyzing, but that
this effect is rather marginal.

We should also mention that it is not possible to compare our results with the proper motions
obtained by Eisl\"offel et al. (1994) and by Bally et al. (2002),
because these authors calculated offsets for small angular
scale structures within the larger scale knots of Herbig \& Jones (1981). It is not
possible to recognize these smaller scale structures in the more recent HST frames that
we are analyzing, as a substantial morphological evolution has occurred (see the discussion
of section 3).

\section{The H$\alpha$ fluxes and the [S~II]/H$\alpha$ line ratios}

From paraboloidal fits to the peaks of the convolutions of the H$\alpha$ and [S~II] images
with a $1''$ wavelet (see the discussion of Section 4 and Figures 5 and 6), we obtain
a peak intensity $I_p$ for each knot. We can then calculate the flux associated with the
peak of the structure as $F_p=\pi \sigma^2 I_p$ (where $\sigma=1''$ is the radius of the
central peak of the wavelet, see Section 4).

{ The H$\alpha$ fluxes and the [S~II]/H$\alpha$ ratios for the HH~2 knots in the four epochs
  are shown in Figures 9 and 10. In these Figures, we show the errors for the H$\alpha$
  fluxes and for the [S~II]/H$\alpha$ line ratios corresponding to the errors
  discussed in section 2.}

Several interesting features can be seen in these plots:
\begin{itemize}
\item knot H (which dominates the emission of HH~2) has an H$\alpha$ flux that has grown
  by a factor of $\approx 2.9$ between 1994 and 2014, while keeping a [S~II]/H$\alpha$ ratio
  with variations of only $\sim 20$\%,
\item knot A shows a decrease by a factor of $\approx 14$ in its H$\alpha$ flux, and an
  increase in its [S~II]/H$\alpha$ ratio from $\approx 0.2$ (in 1994) to $\approx 0.9$
  (in 2014),
\item knot D shows a decrease by a factor of $\approx 1.7$ in H$\alpha$, and an approximately
  constant [S~II]/H$\alpha$ ratio,
\item knot K1 shows an increase by a factor of $\approx 1.6$ in H$\alpha$, and an
  increase by a factor of $\approx 1.3$ in the [S~II]/H$\alpha$ ratio and knot K2
  shows an H$\alpha$ increase of $\approx 2.4$ and
  a [S~II]/H$\alpha$ increase of $\approx 1.8$,
\item knot I shows an almost constant H$\alpha$ flux, and a [S~II]/H$\alpha$ ratio
  with relatively strong variations. This latter feature again might be a result of
  the inconsistent results that we obtain in the H$\alpha$ and [S~II] structures
  for this knot (see Section 3),
\item the remaining knots (L, E and T) show H$\alpha$ fluxes and [S~II]/H$\alpha$
  ratios with only small variations in the period covered by the observations.
\end{itemize}

The strongest H$\alpha$ variabilities are seen in knots H and A. The implications
of these two behaviours will be discussed in the following section.

\section{Interpretation}

\subsection{The dispersion of the proper motions}

Raga et al. (1997) considered the proper motions of a system of condensations that form part
of a single, bow-shock like structure of velocity $v_{bs}$ travelling into an environment
which is moving away from the source at a uniform velocity $v_a$ ($<v_{bs}$).
They show that the total range of possible proper motion velocities along ($\Delta v_x$) and across
($\Delta v_y$) the outflow axis are both equal to $v_{bs}-v_a$. This result is somewhat
counter-intuitive, but is indeed the result obtained for a curved bow shock covering all
possible angles between the shock surface and the outflow axis (similar results, but for the
radial velocities, were obtained previously by Hartigan et al. 1997).
However, because one has a finite (and rather small) number of condensations, the actual values
of $\Delta v_x$ and $\Delta v_y$ can be somewhat smaller than $v_{bs}-v_a$. Also, the proper motions
along the outflow axis have $v_a$ (the outflowing velocity of the pre-bow shock
environment) as a minimum possible value.

Considering the H$\alpha$ proper motion velocities of Table 1 (and not considering
knot I, which has inconsistent H$\alpha$ and [S~II] proper motions), we obtain
$\Delta v_x=167$ and $\Delta v_y=94$~km~s$^{-1}$. Also, the lowest axial proper motion
leads to a $v_a\sim 39$~km~s$^{-1}$ value for a possible outflowing motion of the
pre-bow shock environment. These values indicate that a single bow
shock interpretation of HH~2 implies that the object is moving at a velocity of
$\sim 170$~km~s$^{-1}$ into a low velocity medium, with a $v_a\sim 40$~km~s$^{-1}$
maximum possible velocity.

However, the factor of $\sim 1.8$ discrepancy between $\Delta v_x$ and $\Delta v_y$
indicates that the simple, post-bow shock emission model of Raga et al. (1997) is
probably incorrect (as it predicts that $\Delta v_x\sim \Delta v_y$). This could
be due to:
\begin{itemize}
\item the condensations of HH~2 do not correspond to a single bow shock flow travelling
  into an outflowing environment with a single velocity $v_a$ (which
  would not be surprising given the complex morphology of the object!),
\item the condensations could trace a single bow shock, but the emission could
  come not from the immediate post-shock gas but from material which has been
  substantially mixed with the jet material (which would have the effect of
  re-aligning the velocity of the emitting gas to a direction closer to the outflow
  axis).
\end{itemize}

\subsection{Condensations H and A}

These are the condensations with the strongest H$\alpha$ variabilities, and follow
trends of increasing (knot H) and decreasing (knot A) intensities, that can
be traced back to at least 1954, when knot A was the dominant knot of HH~2 (see Herbig 1969).
Nowadays, condensation H dominates the emission from HH~2, and participates in the long-term
trend (from $\sim 1980$ onwards) of increasing intensities and approximately
constant line ratios noted by Raga et al. (2016b) for HH~2.

As discussed by Raga et al. (2016b), an emitting knot with an increasing luminosity and at
the same time almost constant line ratios could correspond to a high density jet
travelling into a region of increasing environmental density (but with smaller values
than the jet density). If this scenario is applicable to HH~2H, the condensation
should have an almost constant velocity (necessary for obtaining the time-independent
line ratios), with a small velocity decrease as a function of time (resulting from
the increasing environmental density). As discussed in Section 4, a small decrease
in the proper motion of condensation H might be seen in the comparison between
our measurements and the ones of Herbig \& Jones (1981).

From 1994 to 2014, condensation A has had a dramatic decrease in its H$\alpha$ intensity
(by a factor of $\sim 14$), accompanied by a strong increase in its [S~II]/H$\alpha$ ratio.
During this period, this  ratio has changed from a value of $\sim 0.2$
to $\sim 0.9$~. Given the fact that Herbig \& Jones
(1981) judged their proper motion of condensation A to be quite uncertain, it is not
clear whether this knot has decelerated, but if we take their proper motion velocity
at face value, it appears that the condensation has mildly accelerated (see Section
4 and Table 1). This combination of decreasing H$\alpha$ intensity, increasing
[S~II]/H$\alpha$ ratio and basically constant (and possibly slightly growing)
velocity cannot be explained with a model of the emission coming from an
angularly unresolved cooling region behind a stationary shock (for the predictions
from such models see, e.g., Hartigan et al. 1987). However, if knot A had slowed
down in a considerable way (which could be possible given the lack of an error
estimation in the knot A proper motion of Herbig \& Jones 1981), the behaviour
of the H$\alpha$ and [S~II] emission of knot A might indeed be consistent with
plane-parallel shock models.

\section{Summary}

We have used four epochs of H$\alpha$ and [S~II] HST images (covering a time baseline
of approximately 20 years) to carry out measurements of the proper motions of the
condensations of HH~2. Because the subarcsecond structure has relatively large morphological
changes (see Section 3 and Figures 1-4), it is necessary to reduce the resolution of the image in order
to have intensity peaks that can clearly be identified in the successive epochs. We
have done this by convolving the images with a wavelet function having a central peak
of radius $\sigma=1''$ (see Section 4 and Figures 5-6).

In these convolved images, we identify 10 knots which can be clearly seen in all
of the four epochs, and for these knots we calculate the proper motions (deduced from
fits to the positions of the peaks in the four epochs, see Section 4, Figures
7-8 and Table 1). We find that all of the knots have motions consistent
with a constant velocity (i.e., an acceleration or deceleration is not seen
in a clear way for any of the knots). Also, we see that
the H$\alpha$ and [S~II] proper motions are consistent for all of the knots
except knot I.

We can compare our proper motion velocities with the ones
carried out by Herbig \& Jones (1981), who measured the velocities of
knots E, G, H, I, A, D and C on photographic plates covering the
period from 1945 to 1980. For knots E, G, H and D, we find slightly
lower proper motion velocities than the ones of Herbig \& Jones (1981),
and a somewhat higher velocity for knot A (see Table 1). It is not clear
that this latter effect is significant, because Herbig \& Jones (1981) discuss
in detail that their proper motions for knot A are strongly affected by
large morphological changes of this knot. For knot I, we find that our
H$\alpha$ proper motion is somewhat higher and our [S~II] proper motion
is substantially lower than the proper motion of Herbig \& Jones (1981).
For knot C (which was measured by Herbig \& Jones 1981) we have
been unable to obtain a proper motion because of its present day
faintness (in H$\alpha$) and its complex morphological evolution.

The conclusion from this comparison between our proper motions and the
ones of Herbig \& Jones (1981) is that we might be observing a deceleration
of the motion of the HH~2 knots, but that this is a small and somewhat
uncertain effect. Taken at face value, this effect could be interpreted
either in terms of dense clumps which are being braked by the
interaction with the surrounding environment, or as the (broken-up) head
of a jet travelling into an ambient medium of increasing density (as a function
of distance from the outflow source).

We have also measured the H$\alpha$ intensities and [S~II]/H$\alpha$ line ratios
(of the knots seen in the convolved images shown in Figures 5 and 6) in the four
observed epochs. We find that during 1994-2014 many of the observed knots have
small variations in both H$\alpha$ and [S~II]/H$\alpha$ (see Figures 10-11).
The knots with the largest H$\alpha$ variability are:
\begin{itemize}
\item knot H: which has an H$\alpha$ increase by a factor of $\approx 2.4$ and an approximately
  constant [S~II]/H$\alpha$ ratio,
\item knot A: which has an H$\alpha$ decrease by a factor of $\approx 14$ and
  an increase by a factor of $\approx 4$ in its [S~II]/H$\alpha$ ratio.
\end{itemize}
As pointed out by Raga et al. (2016b), who studied the angularly integrated emission of
HH~2, the behaviour of knot H could be interpreted as a shock with a time-independent
shock velocity travelling into an environment of increasing density. The behaviour
of knot A, however, defies an interpretation in terms of a steady shock model.

\begin{deluxetable}{lrrrrrrrr}
\tablecaption{HH 2 proper motions\label{prop}}
\tablewidth{0pt}
\tablecolumns{9}
\tablehead{
  \colhead{knot} & \colhead{$v_x$$^a$} & \colhead{$\epsilon(v_x)$$^a$} & \colhead{$v_y$$^a$} &
  \colhead{$\epsilon(v_y)$$^a$} & \colhead{$v_t$$^a$} &
\colhead{$\epsilon(v_t)$$^a$} & \colhead{$v_{HJ}$$^a$} & \colhead{$\epsilon(v_{HJ})$$^a$} }
\startdata
L  &  46.4  &  4.4 &  -4.7 &  20.1 &  46.7 &   7.7 &  & \\
   &  40.3  &  4.3 &  47.3 &  16.6 &  62.1 &  14.9 &  & \\
E  &  27.5  &  7.1 & -47.4 &  21.7 &  54.9 &  20.8 &    55 &  20 \\
   &  30.0  &  4.8 & -41.1 &  20.8 &  50.9 &  19.0 &  & \\
K1 &  45.6  &  2.9 & -45.8 &  17.9 &  64.6 &  15.3 &  & \\
   &  49.2  &  4.5 & -49.0 &  18.1 &  69.5 &  15.7 &  & \\
K2 &  45.6  &  8.4 & -41.1 &  21.1 &  61.4 &  18.7 &  & \\
   &  44.5  &  4.3 & -45.9 &  15.4 &  63.9 &  13.5 &  & \\
G  &  97.3  &  6.9 &  16.7 &  20.3 &  98.7 &  10.8 &   138 &   8 \\
   &  90.3  &  5.8 &  18.0 &  20.1 &  92.1 &  10.6 &  & \\
H  & 206.6  &  5.6 & -46.5 &  18.1 & 211.8 &  10.1 &   217 &  12 \\
   & 183.5  &  7.6 & -53.9 &  17.8 & 191.2 &  12.0 &  & \\
I  & 234.9  &  6.2 &  -8.9 &  30.1 & 235.1 &   8.5 &   214 &  40 \\
   &  62.3  & 12.3 &   5.6 &  24.5 &  62.5 &  14.3 &  & \\
A  & 144.1  &  4.9 &  28.6 &  15.1 & 146.9 &   8.2 &   138 & $\ldots$$^b$ \\
   & 175.5  & 10.5 &  42.8 &  17.9 & 180.6 &  13.5 &  & \\
T  &  66.0  &  4.2 & -40.9 &  19.6 &  77.6 &  14.7 &  & \\
   &  65.7  &  3.9 & -42.2 &  18.8 &  78.1 &  14.3 &  & \\
D  &  39.3  &  4.7 & -12.5 &  19.2 &  41.2 &  11.5 &    77 &  20 \\
   &  32.6  &  6.0 &  -6.1 &  18.9 &  33.1 &  10.0 &  & \\
\enddata
{\baselineskip=0pt
  \tablenotetext{a}{The proper motion velocities and their errors (given in km s$^{-1}$) are calculated assuming a distance
    of 414~pc to HH~2: $v_x$ is the velocity along the outflow axis, $v_y$ across the outflow axis, $v_t$
  is the total proper motion velocity, and $v_{HJ}$ is the proper motion velocity obtained by Herbig \& Jones (1981),}
  \tablenotetext{b}{Herbig \& Jones (1981) do not give an error estimate for their knot A proper motion.}}
\end{deluxetable}

\begin{acknowledgements}
Support for this work was provided by NASA through grant HST-GO-13484 from the
Space Telescope Science Institute. AE, PV and ACR acknowledge support from the CONACyT grants
101356, 101975 and 167611 and the DGAPA-UNAM grants IN109715 and IG100516.
\end{acknowledgements}

\begin{figure}
\centering
\includegraphics[width=15cm]{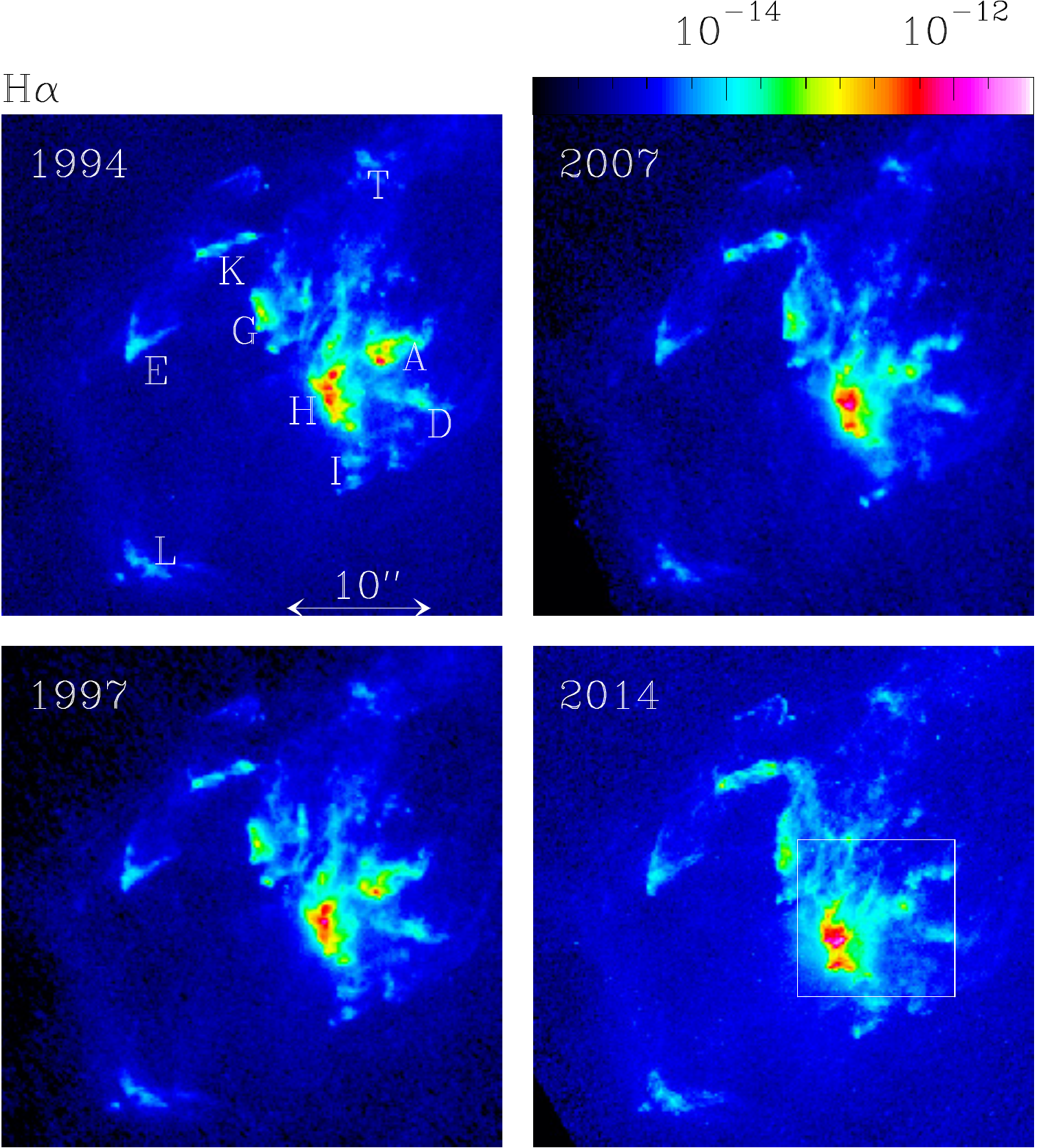}
\caption{The four available epochs of HST H$\alpha$ frames showing a region including all of the
  detected HH~2 emission. The four frames are labeled with the year in which the observations were
  made, and the angular scale is given in the top left frame. The frames are oriented with N on top
  and E to the left. The emission is shown (in erg~s$^{-1}$~cm$^{-2}$~arcsec$^{-2}$) with the logarithmic
  colour scale given by the top right bar. In the top left frame, the knot identifications are given,
  following Herbig \& Jones (1981). In the bottom right frame, the box shows the region which is covered
  in the images shown in Figure 2.}
\end{figure}

\begin{figure}
\centering
\includegraphics[width=15cm]{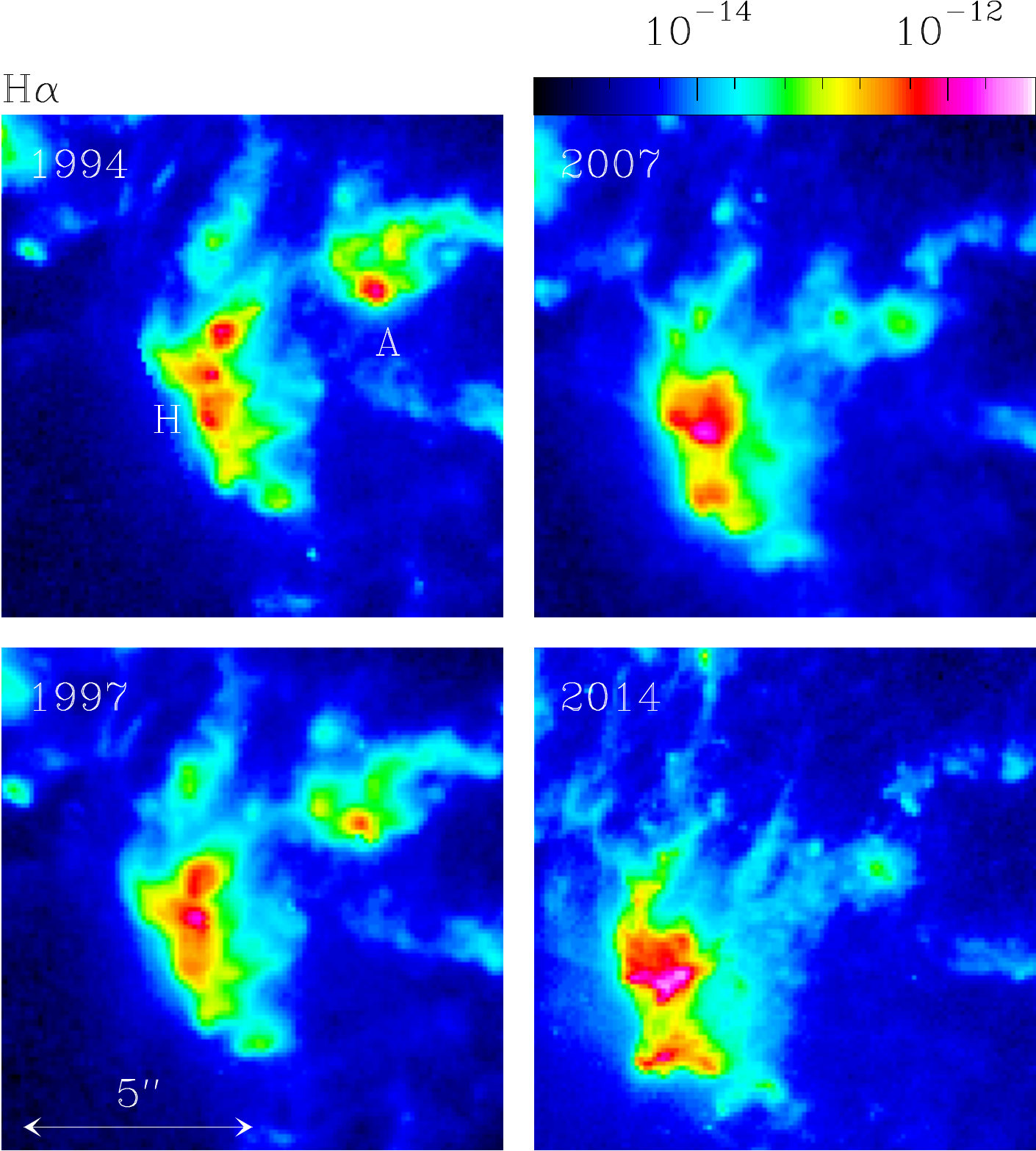}
\caption{This figure shows the H$\alpha$ emission of the central region of HH~2 (defined by the box in the bottom
right frame of Figure 1) in the four epochs of HST images.}
\end{figure}

\begin{figure}
\centering
\includegraphics[width=15cm]{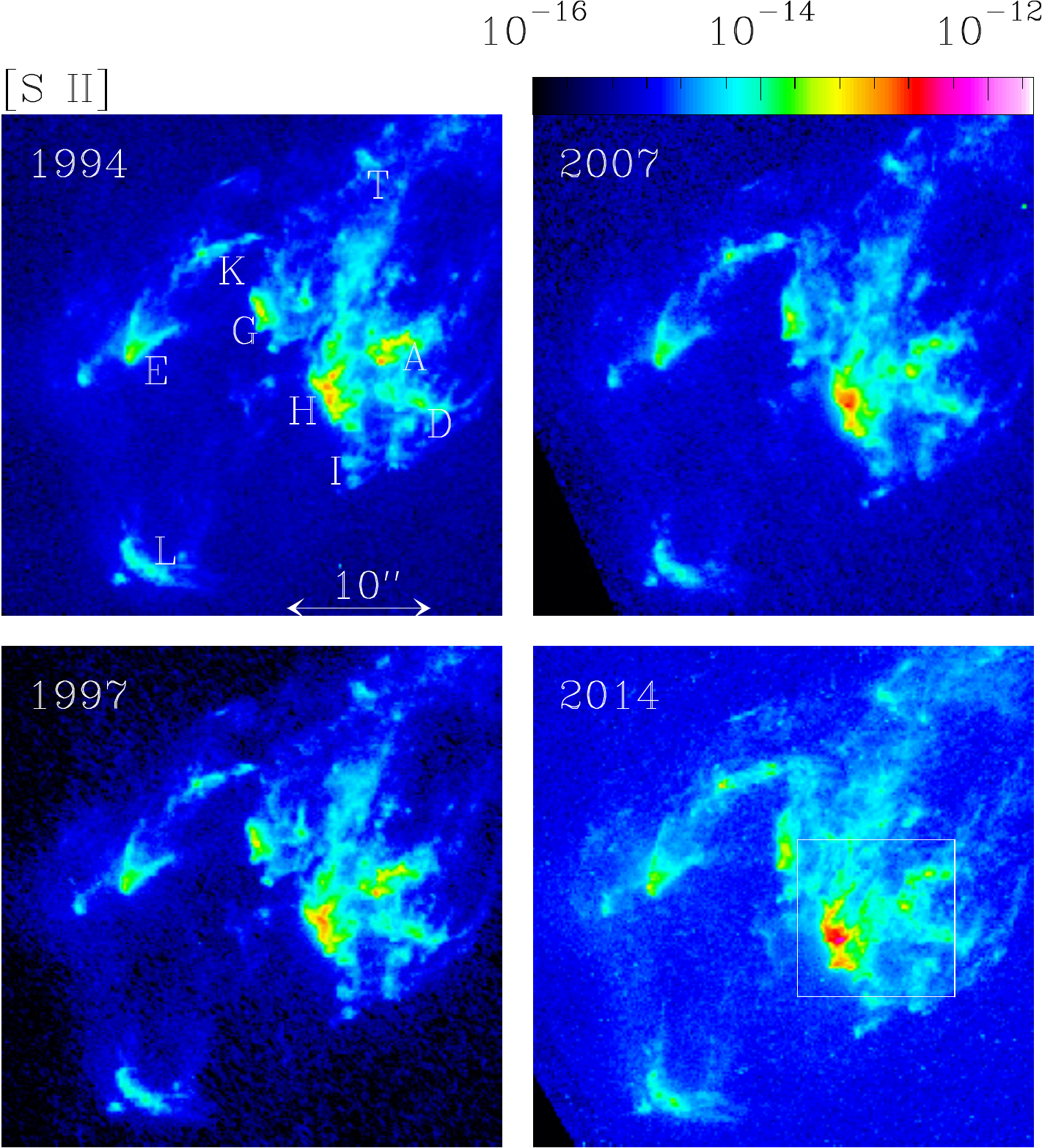}
\caption{The four available epochs of HST [S~II] frames showing a region including all of the
  detected HH~2 emission. This figure has the same characteristics as Figure 1 (which shows
  the corresponding H$\alpha$ frames).}
\end{figure}

\begin{figure}
\centering
\includegraphics[width=15cm]{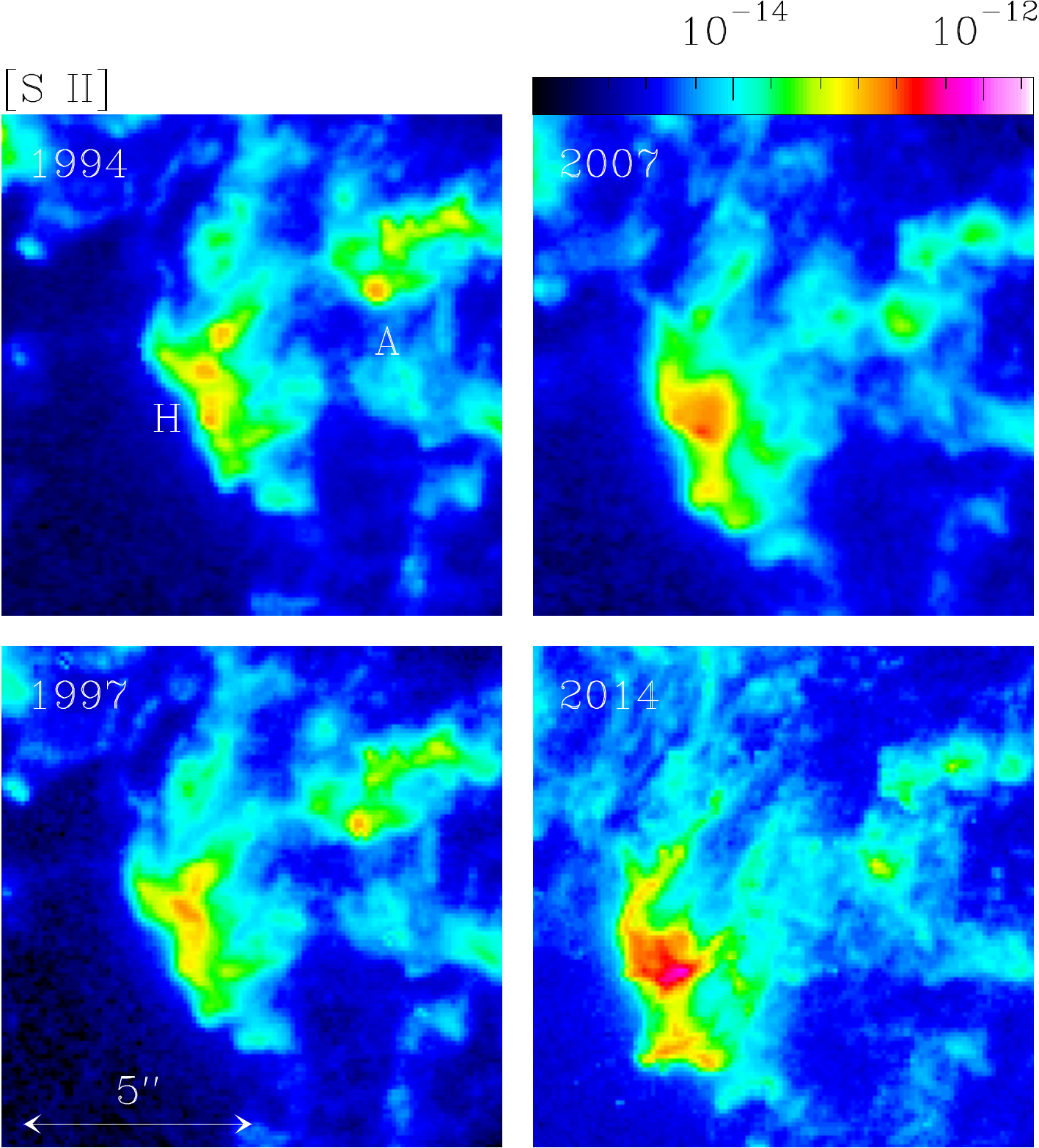}
\caption{This figure shows the [S~II] emission of the central region of HH~2 (defined by the box in the bottom
right frame of Figure 3) in the four epochs of HST images.}
\end{figure}

\begin{figure}
\centering
\includegraphics[width=15cm]{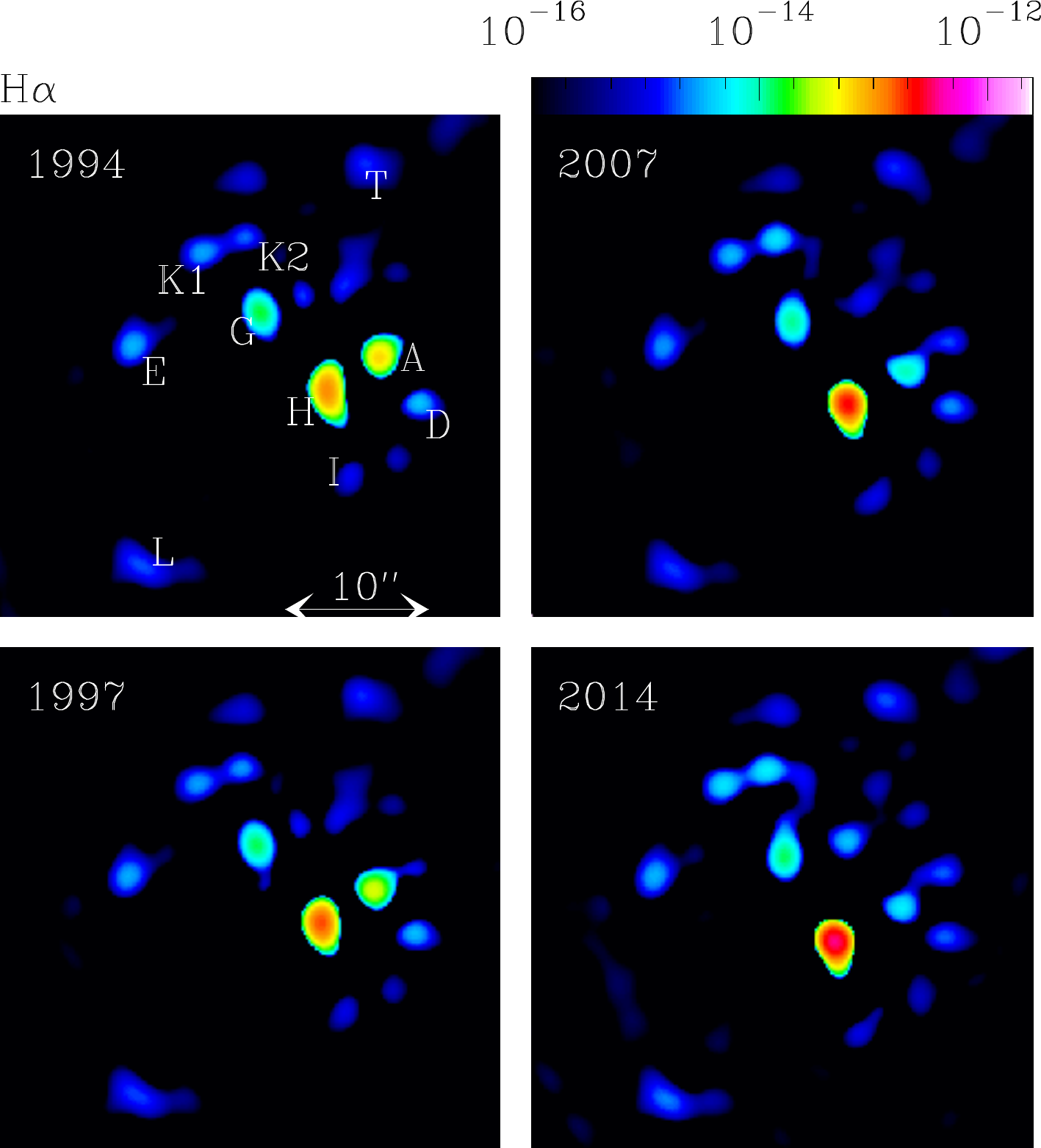}
\caption{Convolutions of the H$\alpha$ images with a $\sigma=1''$ radius ``Mexican hat'' wavelet. The
organization of this figure is the same as the one of Figure 1 (which shows the original H$\alpha$ frames).}
\end{figure}

\begin{figure}
\centering
\includegraphics[width=15cm]{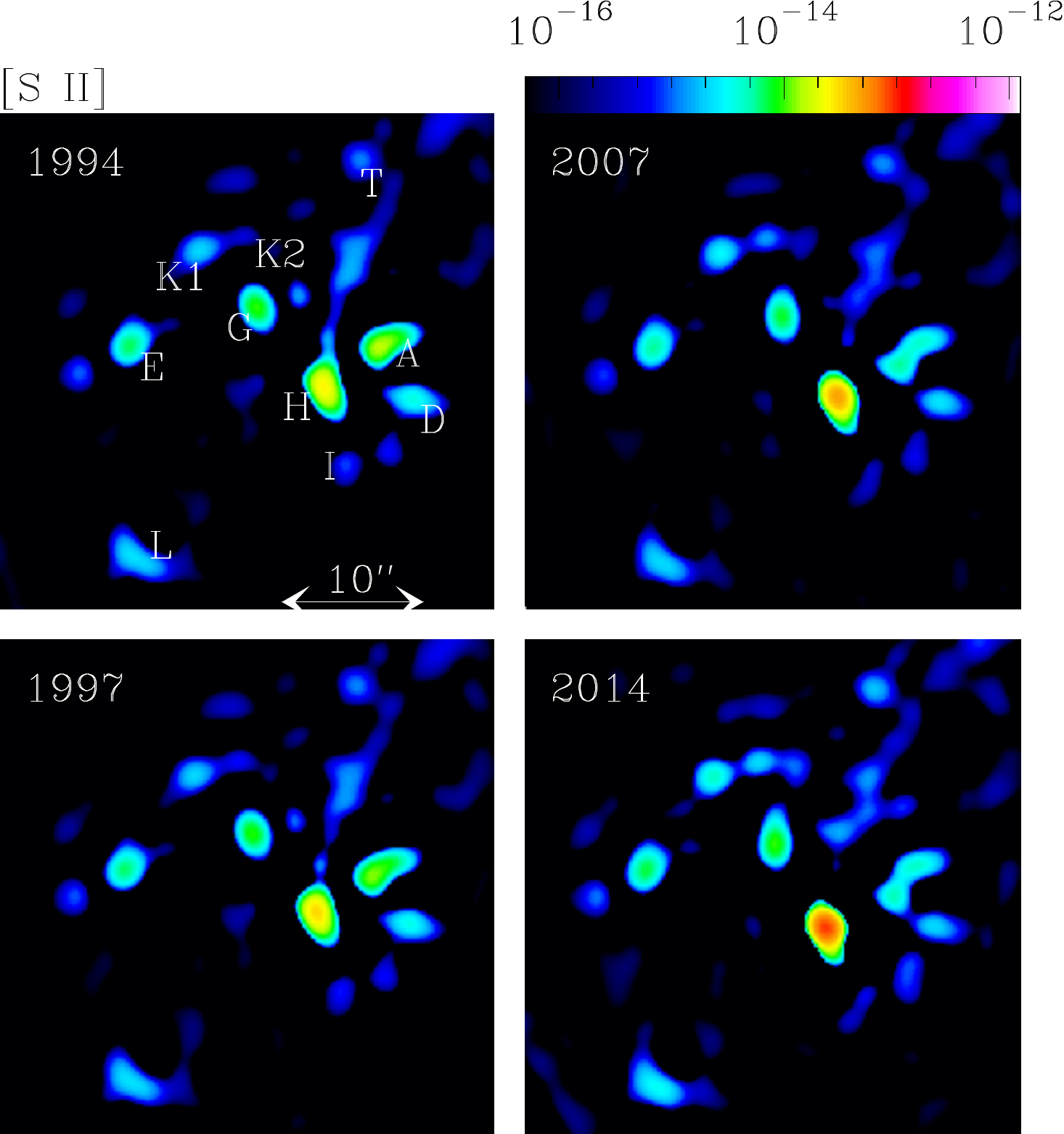}
\caption{Convolutions of the [S~II] images with a $\sigma=1''$ radius ``Mexican hat'' wavelet. The
  organization of this figure is the same as the one of Figure 3 (which shows the original [S~II] frames).}
\end{figure}

\begin{figure}
\centering
\includegraphics[width=15cm]{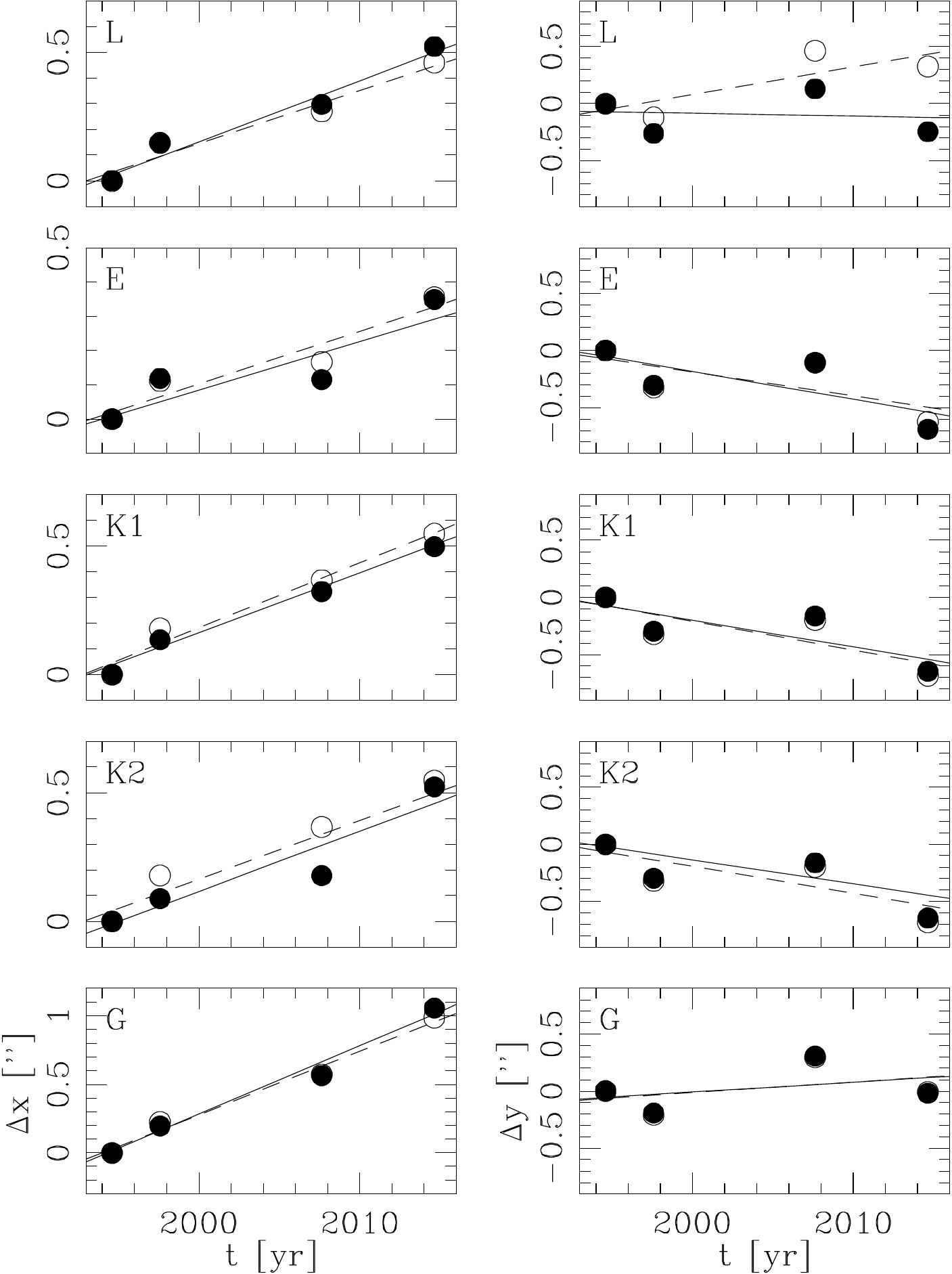}
\caption{Offsets $\Delta x$ (left column, along the outflow axis) and $\Delta y$ (right column, across
  the outflow axis, with positive values directed to the W) from the 1994 positions  as a function of time
  for knots L, E, K1, K2 and G. The H$\alpha$ offsets are shown with filled circles (and linear
  fits to the offsets with solid lines) and the [S~II] offsets with open circle (and linear fits
  with dashed lines).}
\end{figure}

\begin{figure}
\centering
\includegraphics[width=15cm]{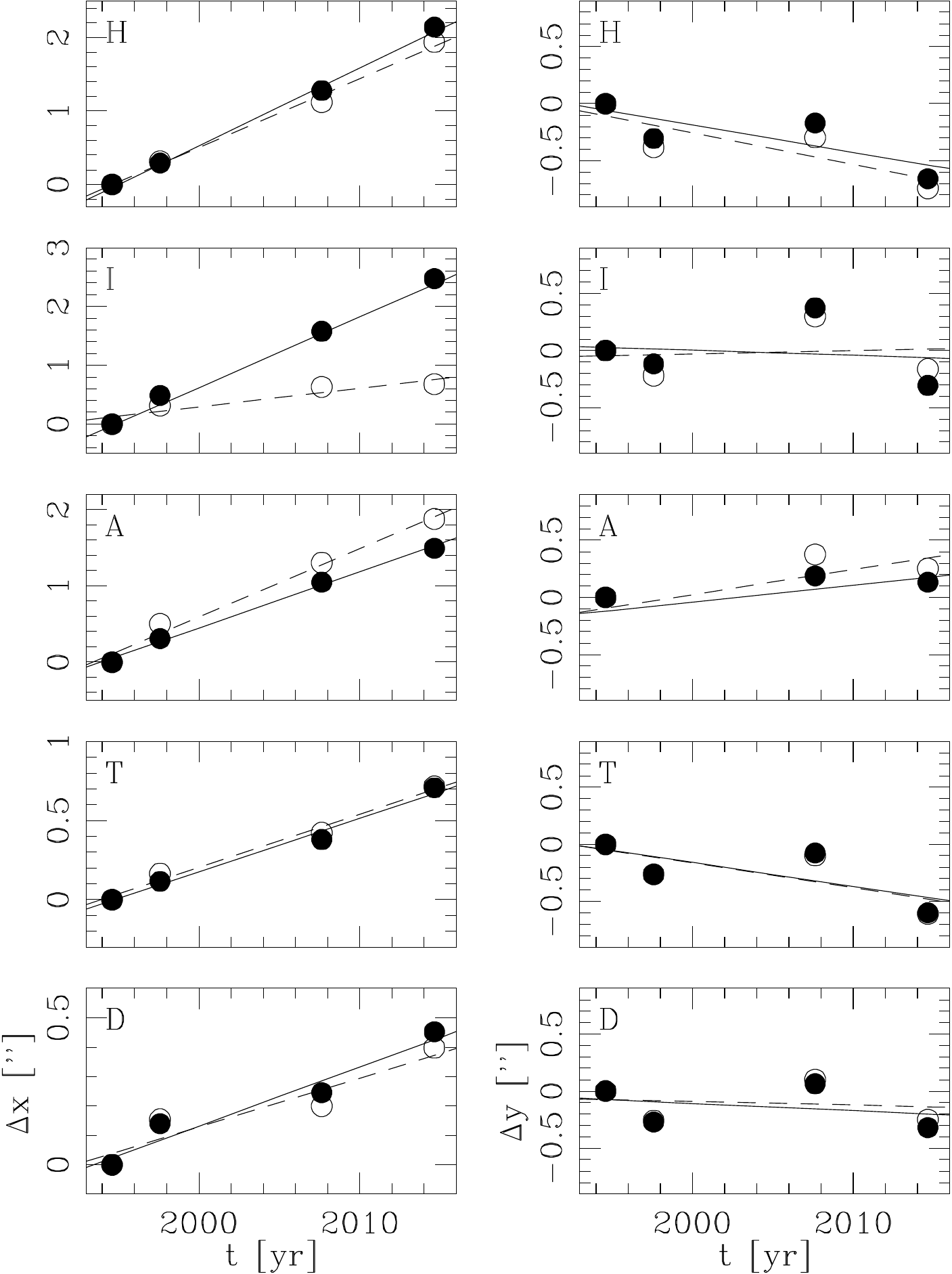}
\caption{Offsets $\Delta x$ (left column, along the outflow axis) and $\Delta y$ (right column, across
  the outflow axis, with positive values directed to the W) from the 1994 positions  as a function of time
  for knots H, I, A, T and D. The H$\alpha$ offsets are shown with filled circles (and linear
  fits to the offsets with solid lines) and the [S~II] offsets with open circle (and linear fits
  with dashed lines).}
\end{figure}

\begin{figure}
\centering
\includegraphics[width=15cm]{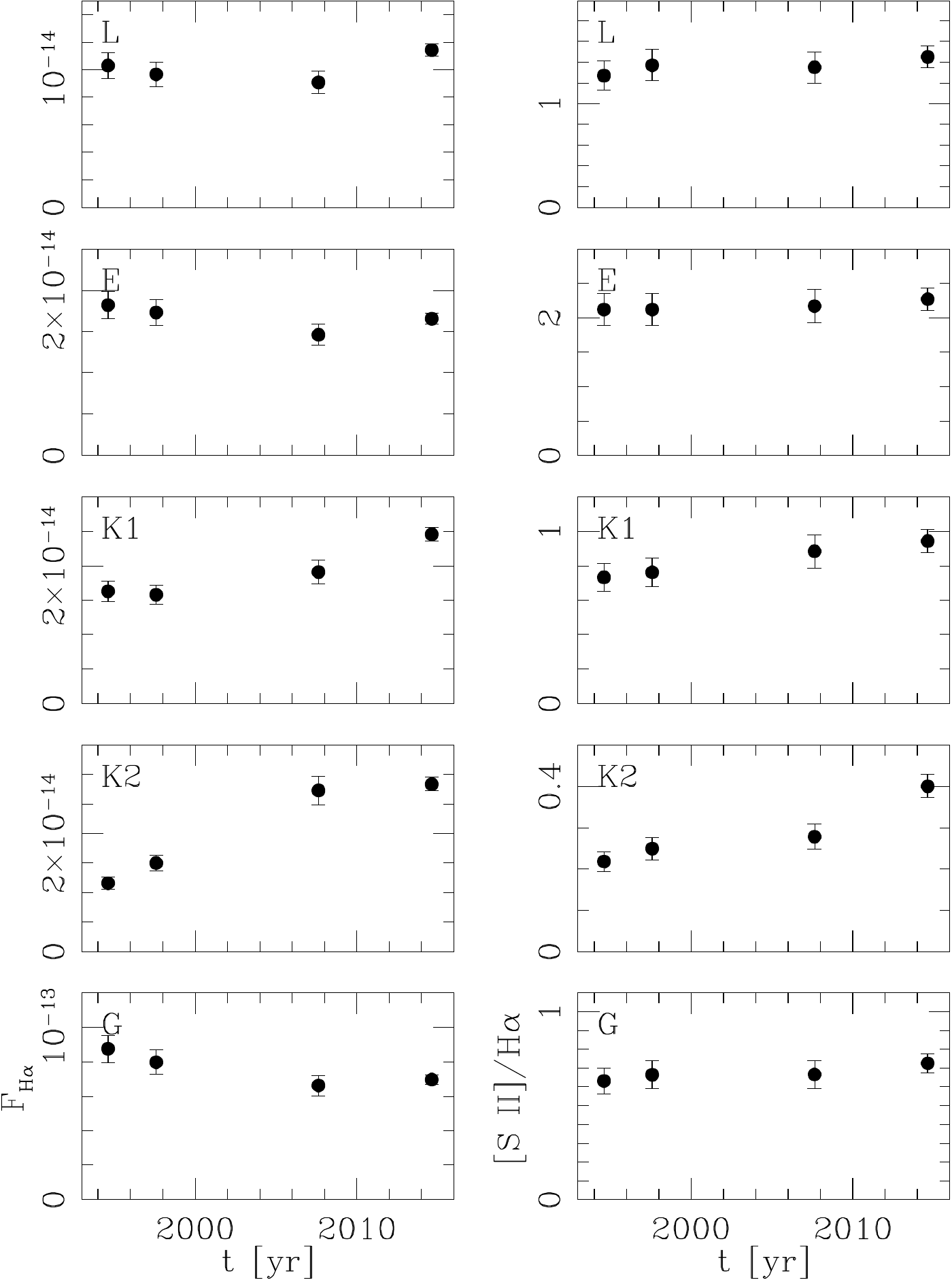}
\caption{H$\alpha$ fluxes (left column, in erg~s$^{-1}$~cm$^{-2}$) and [S~II]/H$\alpha$ ratios (right column)
as a function of time for knots  L, E, K1, K2 and G.}
\end{figure}

\begin{figure}
\centering
\includegraphics[width=15cm]{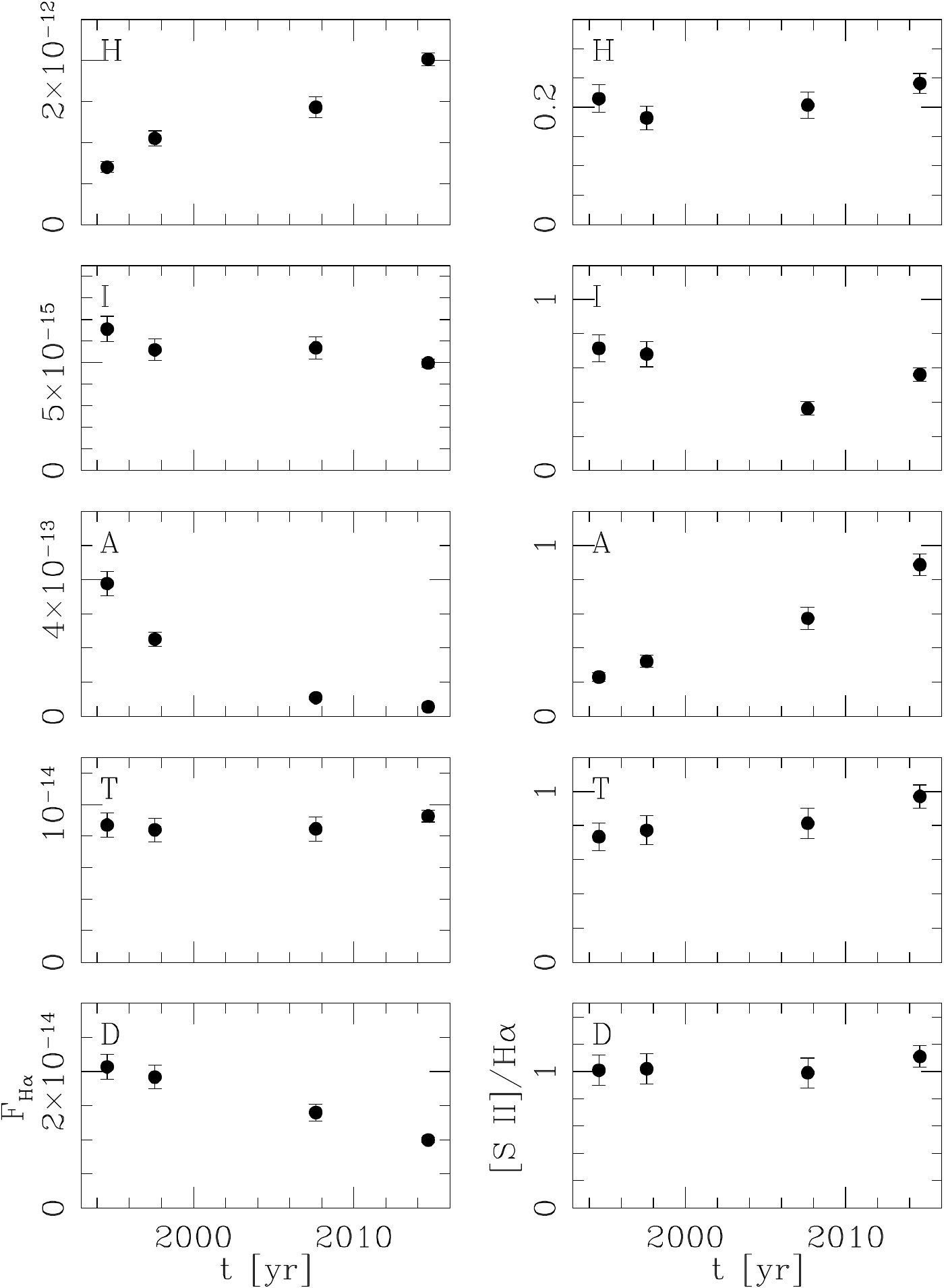}
\caption{H$\alpha$ fluxes (left column, in erg~s$^{-1}$~cm$^{-2}$) and [S~II]/H$\alpha$ ratios (right column)
as a function of time for knots H, I, A, T and D.}
\end{figure}


\begin{thebibliography}

\bibitem[{Bally}{et~al.}(2002)]{bal1} Bally, J., Heathcote, S., Reipurth, B., Morse, J., Hartigan, P.,
  \& Schwartz, R. D. 2002, AJ, 123, 2627

\bibitem[{B\"om} \& {Solf}(1985)]{boh85} B\"ohm, K. H., \& Solf, J. 1985, ApJ, 294, 533

\bibitem[{Brugel}{et~al.}(1981)]{bru1} Brugel, E. W., B\"ohm, K. H., \& Mannery, E. 1981, ApJS, 47, 117

\bibitem[{auth}(year)]{co1} Cohen, M., \& Schwartz, R. D. 1979, ApJ, 233, L77

\bibitem[{auth}(year)]{eis1} Eisl\"offel, J., Mundt, R., \& B\"ohm, K. H. 1994, AJ, 108, 1042

\bibitem[{auth}(year)]{har1} Hartigan, P., Frank, A., Foster, J. M., et al. 2011, ApJ, 736, 29

\bibitem[{auth}(year)]{har2} Hartigan, P., Raymond, J., \& Hartmann, L. W. 1987, ApJ, 316, 323

\bibitem[{auth}(year)]{her1} Herbig, G. H. 1969, Comm. of the Konkoly Obs., No. 65 (Vol VI, 1), p. 75

\bibitem[{author}(year)]{her2} Herbig, G. H., \& Jones, B. F. 1981, AJ, 86, 1232

\bibitem[{author}(year)]{hes1} Hester, J. J., Stapelfeldt, K. R., \& Scowen, P. A. 1998, AJ, 116, 372

\bibitem[{author}(year)]{men07} Menten, K. M., Reid, M. J.,
Forbrich, J., \& Brunthaler, A. 2007, A\&A, 474, 515

\bibitem[{author}(year)]{nor1} Noriega-Crespo, A., \& Raga, A. C. 2012, ApJ, 750, 101

\bibitem[{O'Dell}{et~al.}(2013)]{ode13} O'Dell, C. R., Ferland, G. J., Henney, W. J.,
Peimbert, M. 2013, AJ, 145, 92

\bibitem[{author}(year)]{pra1} Pravdo, S. H., Rodr\'\i guez, L. F., Curiel, S., Cant\'o, J.,
Torrelles, J. M., Becker, R. H., \& Sellgren, K. 1985, ApJ, 293, L35

\bibitem[{author}(year)]{rag1} Raga, A. C., Reipurth, B., Esquivel, A., \& Bally, J. 2016a, AJ, 151, 113

\bibitem[{author}(year)]{rag2} Raga, A. C., Reipurth, B., Castellanos-Ram\'\i rez, A., \& Bally, J. 2016b,
RMxAA, in press

\bibitem[{author}(year)]{rag3} Raga, A. C., Reipurth, B., Castellanos-Ram\'\i rez, A.,
Chiang, H.-F., \& Bally, J. 2015a, ApJ, 798, L1

\bibitem[{author}(year)]{rag4} Raga, A. C., Reipurth, B., Castellanos-Ram\'\i rez, A., Chiang, H.-F., \& Bally, J.
2015b, AJ, 150, 105

\bibitem[{author}(year)]{rag5} Raga, A. C., Reipurth, B., Cant\'o, J., Sierra-Flores, M. M., \& Guzm\'an, M. V.
2011, RMxAA, 47, 425

\bibitem[{author}(year)]{rag6} Raga, A. C., Cant\'o, J., Curiel, S., Noriega-Crespo, A., \& Raymond, J. C.
1997, RMxAA, 33, 157

\bibitem[{author}(year)]{rag7} Raga, A. C., Barnes, P. J., \& Mateo, M. 1990, AJ, 99, 1912

\bibitem[{author}(year)]{rod1} Rodr\'\i guez, L. F., Delgado-Arellano, V. G., G\'omez, Y., et al. 2000, AJ, 119, 882

\bibitem[{author}(year)]{rod2} Strom, S. E., Strom, K. M., Grasdalen, G. L. et al. 1985, AJ, 90, 2281

\end{thebibliography}
\end{document}